\documentclass[10pt,superscriptaddress,amsmath,amssymb,aps,prx,twocolumn,a4paper,floatfix,longbibliography]{revtex4-1}

\usepackage[version=3]{mhchem}
\usepackage{graphicx,times,bm,amssymb,amsmath,booktabs,natbib,physics,subfigure,makecell,hyperref,multirow,color}
\hypersetup{colorlinks,,allcolors=black}
\def\be{\begin{equation}}       \def\ee{\end{equation}}
\def\bea{\begin{eqnarray}}      \def\eea{\end{eqnarray}}

\usepackage{graphicx}
\usepackage{epstopdf}

\begin{document}
\title{\ce{BaCuS2}: a superconductor with moderate electron-electron correlation }

\author{Yuhao Gu}
\affiliation{Beijing National Laboratory for Condensed Matter Physics,
and Institute of Physics, Chinese Academy of Sciences, Beijing 100190, China}

\author{Xianxin Wu}
\affiliation{Beijing National Laboratory for Condensed Matter Physics,
and Institute of Physics, Chinese Academy of Sciences, Beijing 100190, China}

\author{Kun Jiang}
\affiliation{Beijing National Laboratory for Condensed Matter Physics,
and Institute of Physics, Chinese Academy of Sciences, Beijing 100190, China}

\author{Jiangping Hu}
\email{jphu@iphy.ac.cn}
\affiliation{Beijing National Laboratory for Condensed Matter Physics,
and Institute of Physics, Chinese Academy of Sciences, Beijing 100190, China}
\affiliation{CAS Center of Excellence in Topological Quantum Computation and Kavli Institute of Theoretical Sciences,
	University of Chinese Academy of Sciences, Beijing 100190, China}

\begin{abstract}
 We show that the layered-structure \ce{BaCuS2} is a moderately correlated electron system in which the electronic structure of the CuS layer  bears a resemblance to those in both cuprates and iron-based superconductors.  Theoretical calculations reveal that the in-plane $d$-$p$  $\sigma^*$-bonding bands are isolated near the Fermi level. As the energy separation between the $d$ and $p$ orbitals are much smaller than those in cuprates and iron-based superconductors, \ce{BaCuS2} is expected to be moderately correlated. We suggest that this material is an ideal system to study the  competitive/collaborative nature between two distinct superconducting pairing mechanisms, namely the conventional BCS electron-phonon interaction and the electron-electron correlation, which may be helpful to  establish the elusive mechanism of unconventional high-temperature superconductivity.

\end{abstract}


\maketitle

\section{Introduction}
 From conventional BCS superconductors to unconventional high  $T_c$ superconductors, cuprates\cite{lee}, it has been widely suggested that  the superconducting mechanism changes from phonon-mediated attraction to electron-electron correlation driving pairing. However, the existence of such a difference is still under intensive debate. For example, in the search of conventional BCS superconductors with relatively high transition temperature, a possible guiding principle is to find systems with metallized $\sigma$-bonding electrons from light atoms so that the electron-phonon interaction can be maximized\cite{mgb2_pickett,gao2015prediction,gao2015chinese}. Such a textbook example is MgB$_2$\cite{mgb2_mazin,mgb2_pickett,mgb2_xxx}, in which the in-plane $\sigma$ bands are formed by  the p$_x$ and p$_y$ orbitals of B atoms. The Mg$^{2+}$ ions further lower the B $\pi$ ($p_z$) bands, which causes a charge transfer from $\sigma$ to $\pi$ bands and drives the self-doping of the $\sigma$ band. The 2D nature of $\sigma$ bonds then leads to an extremely large deformation potential for the in-plane $E_{2g}$ phonon mode, which greatly enhances the electron-phonon coupling \cite{mgb2_xxx,mgb2_pickett}.  This principle was also argued to be valid even for cuprates, in which the $d$-$p$ $\sigma$-bonding band is responsible for superconductivity \cite{zhang_Rice1988,gao2015chinese}. The argument has left a room to discuss the likelihood of the electron-phonon mechanism in cuprates\cite{zhong2016nodeless}.

  However, the $d$-$p$ bonding displays fundamental differences  from the  $p$-$p$ bonding because of the multiplicity and strong localization of the $d$-orbitals. Such a simple extension is highly questionable. For example, the absence of clear isotope effect \cite{franck1994physical} ,  unconventional  electronic properties in normal states and  strongly antiferromagnetic fluctuations in cuprates suggest the electron-electron correlation can be responsible for high $T_c$  superconductivity\cite{millis1990phenomenological,scalapino1995dwave}. Furthermore, the discovery of high $T_c$ iron-based superconductors(SCs) finishes another large part of the superconducting jigsaw puzzle. Much evidence has suggested that the superconductivity of iron-based superconductors originates from the Fe-As/Te plane with strong electron-electron correlation and is very similar to cuprates\cite{dai2012magnetism,seo2008pairing,si2008strong}. On the other hand, electron-phonon coupling also plays a non-negligible role in iron-based superconductors\cite{Fe_2009nature_isotope,Fe_replicaband_2017coexistence,Fe_2020ultrafast}.

Recently,  focusing on the $d$-orbitals, emphasizing electron-electron interaction, we have suggested  a new guiding principle    for the search of unconventional high $T_c$ superconductors: those
$d$-orbitals with $d$-$p$ $\sigma$-bondings must be isolated near Fermi energy. Under this principle, local cation complexes, the connection between the complexes, the electron filling factor at transition metal atoms and lattice symmetries must collaborate to fulfill the criteria\cite{hu2015predicting,hu_identifying_2016,le2018possible,hu2018stannite,gu2019ni}. This simple principle can explain why cuprates and iron-based superconductors are so special as high  $T_c$ superconductors.

It is noticeable that the above mentioned two principles are linked. While they emphasize different interactions, both of them are featured by $\sigma$-bonding. Thus, why are the two types of  $\sigma$-bonding  fundamentally different? In order to answer these two questions, we want to find a system with  a moderate electron-electron correlation from the $d$-$p$ $\sigma$-bonding  so that an explicit comparison between the electron-electron correlation and electron-phonon interaction can be examined.

 In this paper,  we propose that  a new material BaCuS$_2$ can fulfill the above task. \ce{BaCuS2} is a moderately correlated electron system in which the $d$-$p$  $\sigma^*$-bonding bands solely control the electronic physics near Fermi energy. Similar to cuprates, iron pnictides and MgB$_2$, BaCuS$_2$ also has a layered structure, where the electronic structure is dominated by the CuS$_2$ square layer.  We demonstrate that the electron-electron correlation may drive superconductivity in \ce{BaCuS2} with a $d_{xy}$-wave pairing symmetry, very similar to the superconductivity in cuprates. While, if the superconductivity is caused by electron-phonon couplings, a conventional BCS $s$-wave state is expected,  similar to \ce{La3N3Ni2B2}. According to first-principles calculations, we find that the BaCuS$_2$ phase is  thermodynamically stable and has lower formation energy under pressure compared with other known phases, suggesting that it can be synthesized in future experiments under external pressure.

\section{B\lowercase{a}C\lowercase{u}S$_2$ electronic structure and Comparison with other SC materials}

\begin{figure}[htb]
	\centerline{\includegraphics[width=0.5\textwidth]{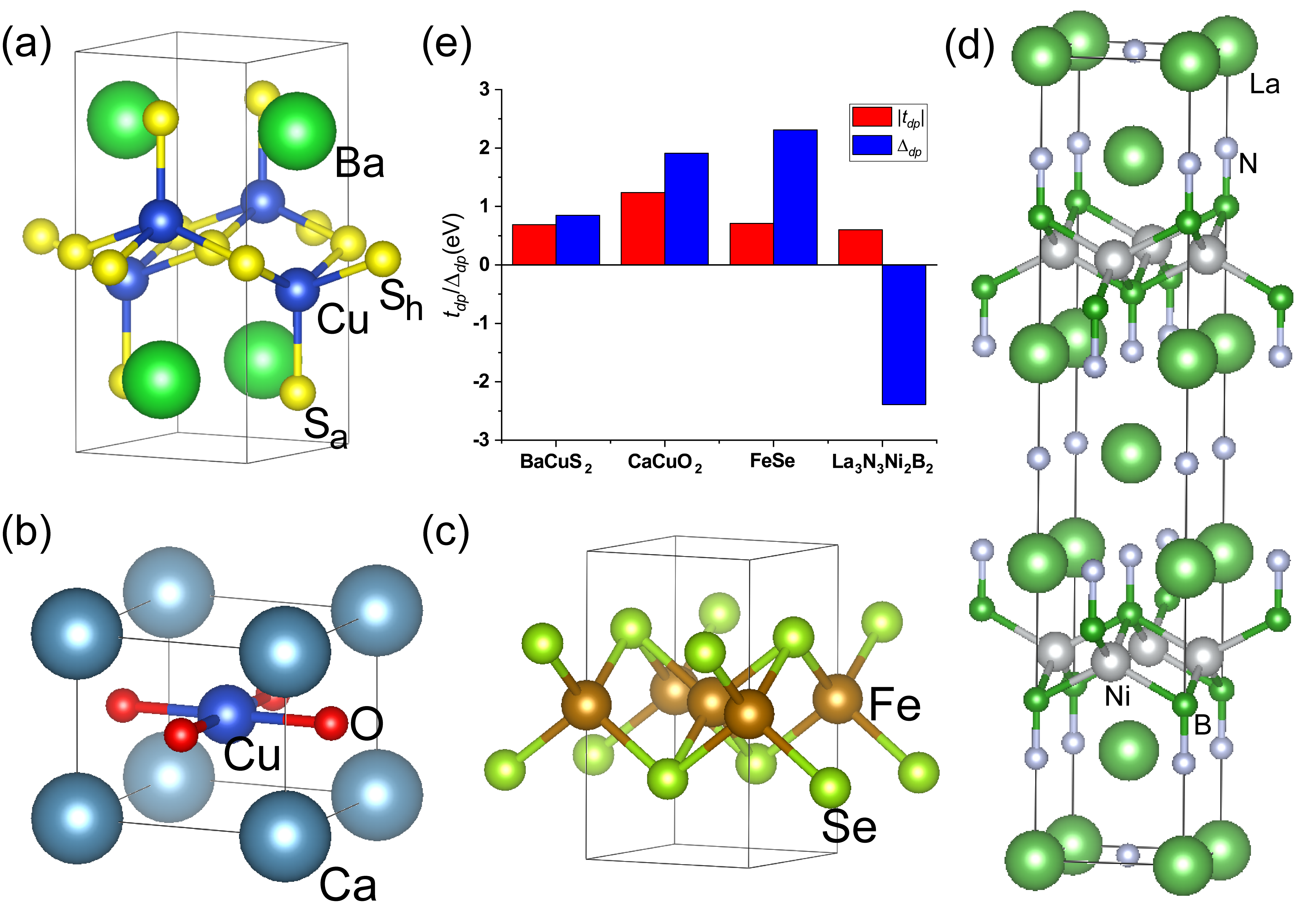}} \caption{(a) Crystal structure of \ce{BaCuS2}. Here S$_a$ represents the apical S atoms while S$_h$ represents the horizontal S atoms. (b-d) Crystal structures of other layered superconductors (b) \ce{CaCuO2}, (c) \ce{FeSe} and (d) \ce{La3N3Ni2B2}. (e) The comparison of parameters in $dp$-models of \ce{BaCuS2}, \ce{CaCuO2}, \ce{FeSe} and \ce{La3N3Ni2B2}. Here $|t_{dp}|$ is the amplitude of the major hopping parameter in each compound ($|t_{{Cu,d_{x^2-y^2}}-{S_h,p_{x/y}}}|$ in \ce{BaCuS2}, $|t_{{Cu,d_{x^2-y^2}}-{O,p_{x/y}}}|$  in \ce{CaCuO2},
		$|t_{{Fe,d_{xz/yz}}-{Se,p_{x/y}}}|$ in \ce{FeSe} and $|t_{{Ni,d_{xz/yz}}-{B,p_{x/y}}}|$  in \ce{La3N3Ni2B2}) and $\Delta_{dp}$ is the corresponding on-site energy difference.
		\label{fig1} }
\end{figure}

\begin{figure}[htb]
	\centerline{\includegraphics[width=0.5\textwidth]{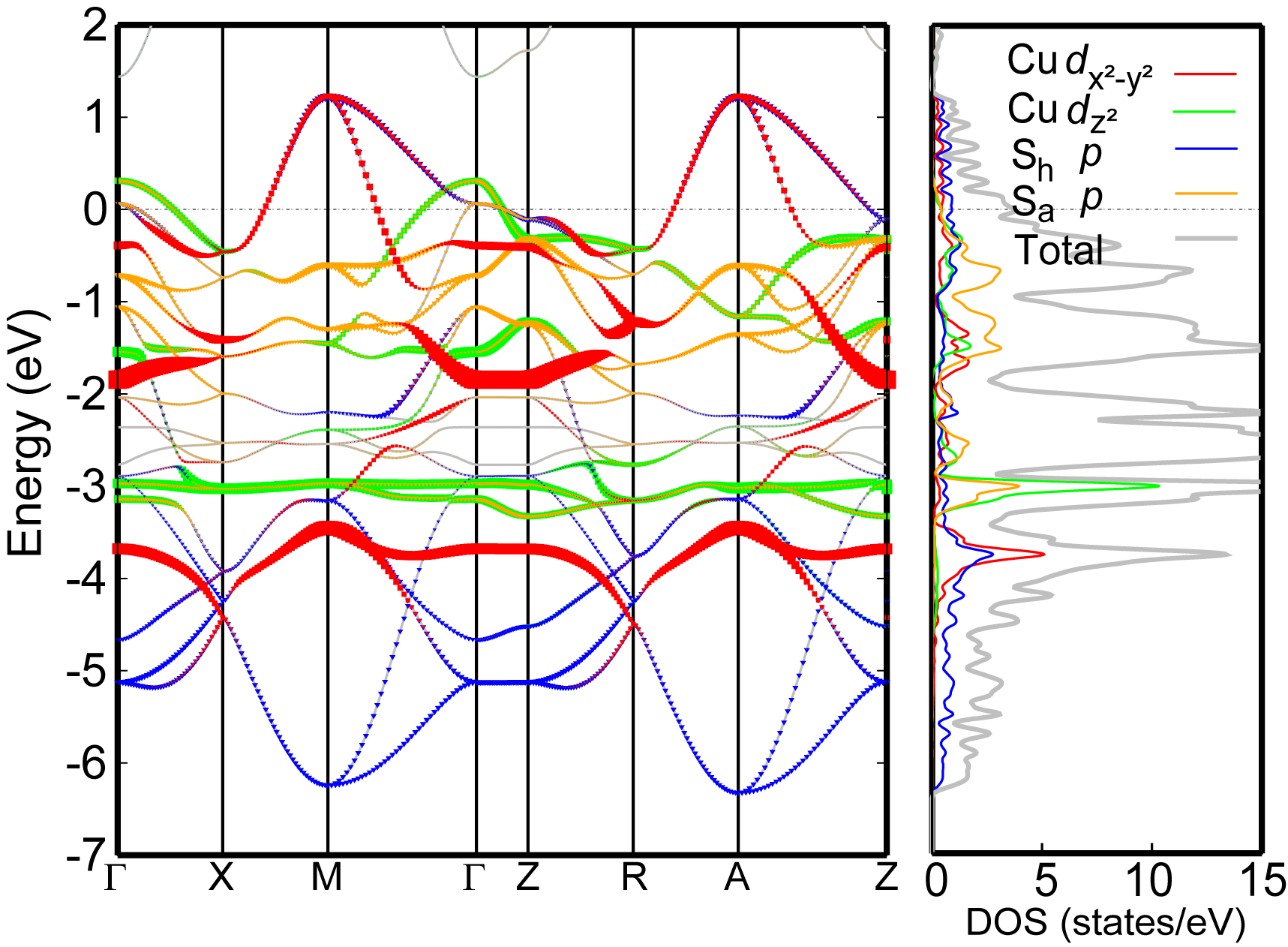}} \caption{The band structure and density of states for \ce{BaCuS2} from density functional theory (DFT) calculation. The sizes of dots represent the weights of the projection. Here S$_a$ represents the apical S atoms while S$_h$ represents the horizontal S atoms.
		\label{fig2} }
\end{figure}

We start from the crystal structure and electronic structure. The layered ternary transition metal sulfide \ce{BaCuS2} has a structure: the BaS layers alternate with \ce{Cu2S2} layers and the transitional metal atom is in square pyramidal coordination, as shown in FIG.\ref{fig1}(a). The upward-pointing square pyramidals connect the downward-pointing ones by sharing edges, forming the glide symmetric \ce{Cu2S2} plane. Similar to other layered SCs, the main electronic structure of \ce{BaCuS2} stems from its \ce{Cu2S2} layer.  To demonstrate it, we carried out density functional theory calculations for \ce{BaCuS2}. The electronic structure and density of states (DOS) of \ce{BaCuS2} are plotted in FIG.\ref{fig2}. The bands around the Fermi level are formed by the $d$-$p$ valence manifold. Partially-filled $d$-$p$  $\sigma^*$-bonding bands cross the Fermi level, where Cu $d_{x^2-y^2}$ orbitals strongly hybridize with in-plane S $p_{x/y}$ orbitals and Cu $d_{z^2}$ orbitals strongly couple with apical S $p_z$ orbitals. Owing to the planar nature of Cu $d_{x^2-y^2}$ orbitals and S $p_{x/y}$ orbitals, these bands have a weak dispersion along the $k_z$ direction. In contrast, the bands from Cu $d_{z^2}$ orbitals and apical S $p_z$ orbitals exhibit a large dispersion. We expect that the SC of \ce{BaCuS2} is mainly contributed from the 2D cylindrical Fermi surface, which is similar to the other layered SCs.



It is interesting to compare the electronic structure of \ce{BaCuS2} with those of high-$T_c$ superconductors (\ce{CaCuO2}\cite{azuma1992CaCuO2110K}; \ce{FeSe}\cite{hsu2008FeSe8K}) and known BCS superconductor (\ce{La3N3Ni2B2}\cite{cava1994superconductivity}). These three materials are layered transition metal compounds, are shown in FIG.\ref{fig1}(b-d). Their electronic structures are mainly attributed to the square layers.  In \ce{CaCuO2} and \ce{FeSe}, the electron correlation plays an important role in the unconventional superconductivity\cite{hu_identifying_2016}. From an electronic-structure perspective, this is consistent with the fact that the $d$-orbitals dominate around the Fermi surfaces in these materials, giving rise to strong correlations. However, in \ce{La3N3Ni2B2}, the extended $s$-$p$ bands of anions dominate Fermi surfaces and are featured by strong electron-phonon couplings via B's high frequency $A_{1g}$ phonons, while Ni's $d$-orbitals play a less pronounced role\cite{mattheiss1995La3Ni2B2N3,Pickett1995La3N3Ni2B2}.  As shown in FIG.\ref{fig2}, in the $d$-$p$ $\sigma^*$-bonding bands of \ce{BaCuS2} near Fermi energy, the weight of $p$-orbitals of the in-plane S atoms is much larger than those in cuprates and iron-based superconductors. In contrast to \ce{La3N3Ni2B2},  the weight of $d_{x^2-y^2}$ orbitals of Cu atoms is still sizable.

To quantitatively confirm this point, we construct tight-binding models including the $d$-orbitals of transition-metal atoms and the $p$-orbitals of coordinated anions to analyze their electronic structures by calculating the maximally localized Wannier functions (MLWFs)\cite{see_appendix}. Our Wannierization results successfully reproduce the band structures from DFT calculationsp, as shown in appendix (FIG.\ref{dp_wan}). Then, we extract hopping parameters and on-site energies from our Wannierization results and display the representative parameters in appendix (TABLE.\ref{hops}). In cuprate \ce{CaCuO2} and iron-based superconductor \ce{FeSe}, the electronic physics is dominated by $d$-orbitals near the Fermi level, whose on-site energies are much higher (about 2 eV) that those of coupled $p$-orbitals. Nevertheless, in BCS-type superconductor \ce{La3N3Ni2B2}, where the NiB layer is isostructural to the FeSe layer, the electronic physics is quite distinct: the on-site energy of B-$p_{x/y}$ orbital is even higher than that of Ni-$d_{xz/yz}$ orbital. This is consistent with previous studies\cite{mattheiss1995La3Ni2B2N3,Pickett1995La3N3Ni2B2}: multiple components cross the Fermi level, showing that \ce{La3N3Ni2B2} is a good metal. The scenario of \ce{BaCuS2} is different from above examples: the on-site energies of $d$-orbitals are still higher than the $p$-orbitals, but the energy difference is only about 1 eV, much less than those in high-$T_c$ superconductors. It suggests that the antiferromagnetic (AFM) order may not be stabilized in \ce{BaCuS2} due to the weak superexchange coupling \cite{NOAFM}. Moreover, the difference in hopping parameters between partially-filled $d$-orbitals and coupled $p$-orbital is not so significant, as shown in FIG.\ref{fig1}.(b).

The above analysis in  \ce{BaCuS2} is consistent with the absence of any magnetically ordered states in our calculation. Therefore, \ce{BaCuS2} tends to be a material  with a moderate electron-electron correlation.

 As mentioned previously, the guiding principle for searching high transition temperature BCS superconductors are light atoms and metallized $\sigma$-bonding electrons\cite{gao2015prediction,gao2015prediction,mgb2_pickett}. In principle, in \ce{BaCuS2},  the $d$-$p$ $\sigma^*$-bonding bands cross the Fermi level, where those metallic $\sigma$-bonding electrons can support the BCS-type superconductivity. Thus, we calculate the electron-phonon coupling (EPC) properties of \ce{BaCuS2}\cite{see_appendix}. The EPC strength $\lambda$ is about 0.59 in \ce{BaCuS2}, which is lower than that in \ce{La3N3Ni2B2} ($\lambda \sim 0.86$, $T_c \sim 13$K\cite{poole1999handbook}) but slightly higher than that in \ce{LaNiBN} ($\lambda \sim 0.52$, $T_c \sim 4.1$K\cite{jung2013LaNiBN}). However, due to the heavy mass of Cu and S atoms, we find that the electron-phonon coupling in \ce{BaCuS2} can only induce induce superconductivity of $T_c$ less than 4 K.\cite{see_appendix}.

\begin{table}[ht]
\caption{\label{str-table}%
The optimized structural parameters for \ce{BaCuS2} (space group $P4/nmm$). Here S$_a$ represents the apical S atoms while S$_h$ represents the horizontal S atoms. }
\begin{ruledtabular}
\begin{tabular}{cccccc}
System  & a(\AA) & c(\AA) & Cu-S$_h$(\AA) & Cu-S$_a$(\AA) & Cu-S$_h$-Cu($^\circ$) \\
 \colrule
 \ce{BaCuS2} & 4.49 & 9.14 & 2.41 & 2.32 & 97.6 \\
\end{tabular}
\end{ruledtabular}
\end{table}

\section{ The effective two band model and RPA results}
\begin{figure}
	\centerline{\includegraphics[width=1.0\columnwidth]{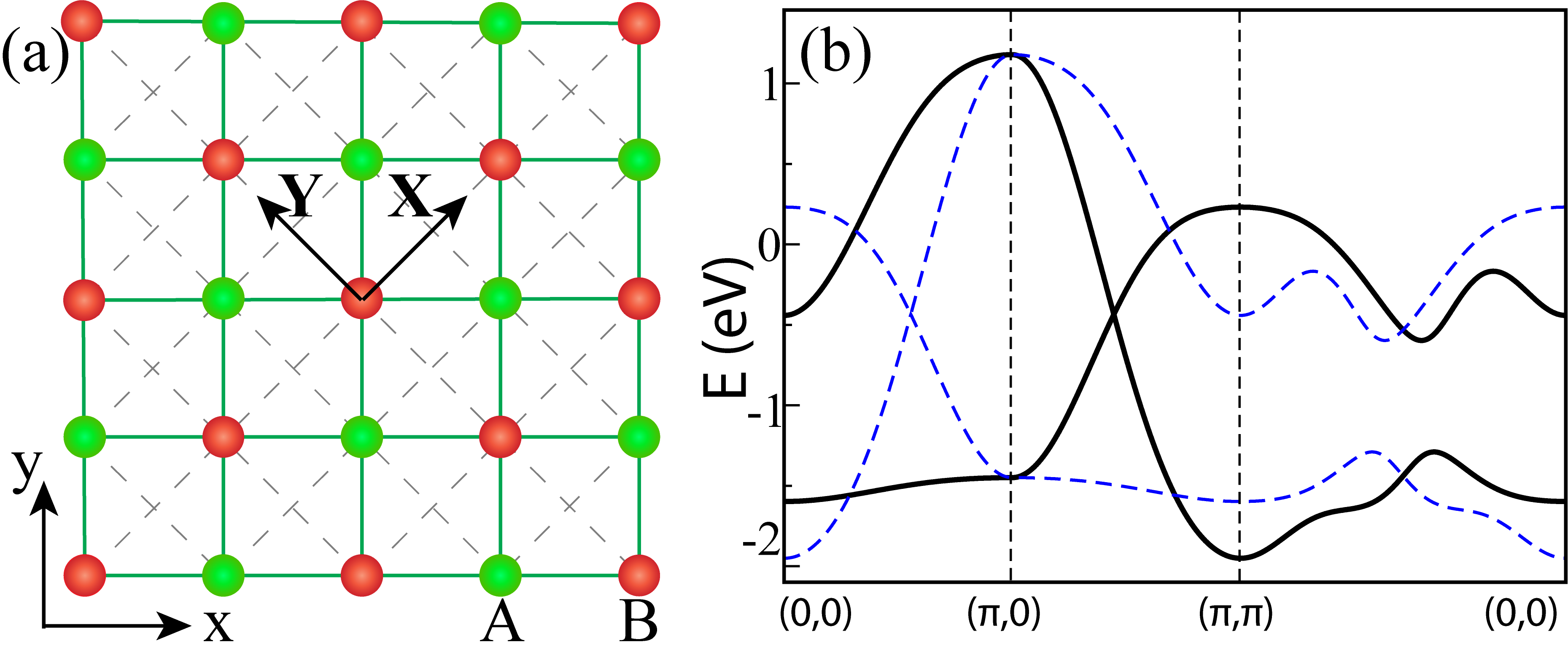}}
	\caption{(color online) (a)  Lattice structure of Cu layer in the tight-binding model. In each unit cell, there are two Cu atoms sitting above and below the $xy$ plane. We label these two sublattices by A and B, respectively. The conventional crystal structure direction is defined for Cu$_A$ to Cu$_A$ direction, labeled as $X-Y$. To simplify our model using glide symmetry, we rotate the global axis to Cu$_A$ to Cu$_B$ direction, labeled as $x$-$y$. (b) The two-Cu tight-binding model for BaCuS$_2$ in Eq.(1) (black lines). The blue dash lines are the folded energy bands with a folding vector Q=$(\pi,\pi)$.}
	\label{tb_band}
\end{figure}
To study the correlation effect, we first construct an effective minimal model  by Wannierization based on the $d_{x^2-y^2}$-like and $d_{z^2}$-like MLWFs on Cu sites in \ce{BaCuS2} \cite{mostofi2008wannier90,Marzari2012,see_appendix}. The $d$-$p$ $\sigma^*$-bonding bands are obviously more delocalized in \ce{BaCuS2} than that in \ce{CaCuO2} (More details can be found in appendix FIG.\ref{sigma*_wan}.(c-e)). As a consequence, the correlation strength in \ce{BaCuS2} should be weaker. Then by fitting the Wannierization results\cite{see_appendix}, we arrive at an effective tight-binding (TB) model in the basis of $d_{x^2-y^2}$ orbital and $d_{z^2}$ orbital.

There are two Cu atoms in each unit cell, as shown in  FIG.\ref{tb_band}(a), which indicates that the minimal model for \ce{BaCuS2} contains four bands. Similar to FeSe, the space group of  \ce{BaCuS2} is $P4/nmm$. There is a glide symmetry which consists of a translation Cu$_A$ to Cu$_B$ and a mirror reflection perpendicular to the S plane. Then, using glide symmetry, one can unfold the band structure into the Brillouin zone of one-Cu unit cell and write down a two-band model for \ce{BaCuS2}. Note that, the conventional crystal structure direction of \ce{BaCuS2} is defined for the Cu$_A$ to Cu$_A$ direction, labeled as $X-Y$. Similar to iron based superconductors, we define a new coordinate system with the $x$ and $y$ axes aligned to the Cu$_A$ to Cu$_B$ direction, labeled as $x$-$y$ \cite{hu2012s,hao2014topological}.

\begin{figure}
	\centerline{\includegraphics[width=1.0\columnwidth]{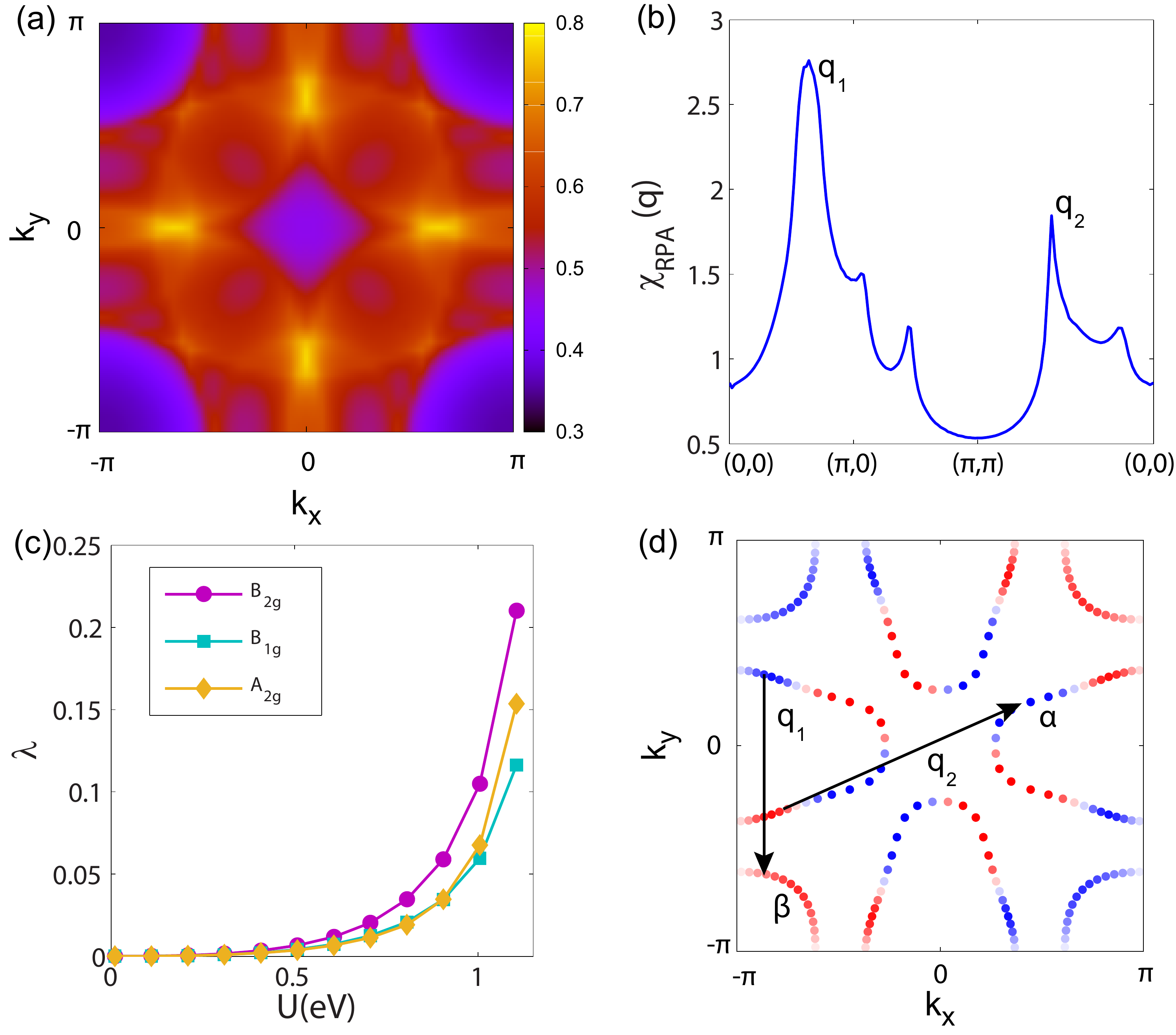}}
	\caption{(color online) (a) Distribution of the largest eigenvalues for bare susceptibility matrices $\chi_0(\bm{k})$ at $n=2.0$. (b) Largest eigenvalues of RPA susceptibility with $U=0.9$ eV and $J/U=0.2$. (c) Pairing strength eigenvalues for the leading states as a function of interaction U with $J/U=0.2$. (d) Gap function of the dominant $B_{2g}$ state ($d_{xy}$-wave pairing).}
	\label{RPA}
\end{figure}

The two-band model in one-Cu unit cell in the basis $\psi_k=(d_{z^2}(k),d_{X^2-Y^2}(k))$ (the spin index is omitted here) can be written as
\begin{eqnarray}
	H=\sum_{k}\psi_k^{\dagger} \hat{H}_k \psi_k
\end{eqnarray}
where the 2 by 2 matrix $H_k$ and more details are provided in appendix. The  band structure in the original unit cell can be obtained by the folding the band structures of the Hamiltonian $H_k$, as plotted in FIG.\ref{tb_band}(b), where the blue dash lines are folded bands with a folding vector Q$=(\pi,\pi)$.  The corresponding FSs are displayed in Fig.\ref{RPA}(d) and the large oval FSs around ($\pi$,0) or (0,$\pi$) are mainly attributed to $d_{X^2-Y^2}$ orbital while the smaller circular FS around the M point is mainly attributed to $d_{z^2}$ orbitals.

 Using the above two-band model, we can apply  the standard approach to investigate the intrinsic spin fluctuations by carrying out RPA calculation\cite{graser2009near,kemper2010sensitivity,xxwu2014g,xxwu2015triplet,xxwu}. We adopt the general multi-orbital Coulomb interactions, including on-site Hubbard intra- and inter-orbital repulsion $U/U^{\prime}$, Hund's coupling $J$ and pair-hopping interactions $J^{\prime}$,
\begin{eqnarray}
H_{i n t}&=& U \sum_{i \alpha} n_{i \alpha \uparrow} n_{i \alpha \downarrow}+U^{\prime} \sum_{i, \alpha<\beta} n_{i \alpha} n_{i \beta} \nonumber \\
&&+J \sum_{i, \alpha<\beta, \sigma \sigma^{\prime}} c_{i \alpha \sigma}^{\dagger} c_{i \beta \sigma^{\prime}}^{\dagger} c_{i \alpha \sigma^{\prime}} c_{i \beta \sigma} \nonumber \\
&&+J^{\prime} \sum_{i, \alpha \neq \beta} c_{i \alpha \uparrow}^{\dagger} c_{i \alpha \downarrow}^{\dagger} c_{i \beta \downarrow} c_{i \beta \uparrow},
\end{eqnarray}
where $n_{i \alpha}=n_{i \alpha \uparrow}+n_{i \alpha \downarrow}$. The distribution of the largest eigenvalues of bare susceptibility matrices is displayed in Fig.\ref{RPA}(a). There is a prominent peak at $q_1$, close to ($\frac{\pi}{2}$,$0$). In addition, the bare susceptibility shows a broad peak at $q_2$. The former is attributed to the inter-pocket nesting between $\alpha$ and $\beta$ and the latter is contributed by intra pocket nesting in $\alpha$ Fermi surface. From the RPA spin susceptibility along high-symmetry lines shown in Fig.\ref{RPA}(b), we find that these peaks get significantly enhanced when interactions are included. All peaks in the susceptibility are far away from the $\Gamma$ point, indicating intrinsic antiferromagnetic fluctuations in the system. To investigate the pairing symmetry, we calculate the pairing strengths as a function of Coulomb interaction $U$ with $J/U=0.2$. The dominant pairing has a $B_{2g}$ symmetry, whose gap function is shown in Fig.\ref{RPA}(d). Each pocket has a $d_{xy}$-wave gap but the intrapocket nesting in $\alpha$ pockets induces additional sign changes in the corresponding gap functions. Moreover, there is a sign change between the gap functions on $\alpha$ and $\beta$ pockets, which is determined by inter pocket nesting. The gap functions on these Fermi surfaces can be qualitatively described by a form factor $sink_xsink_y(cosk_x+cosk_y)$, which is classified as the $d_{xy}$ ($B_{2g}$) pairing symmetry. In this superconducting state, there are gapless nodes on high symmetry lines as well as nodes on the original BZ boundary.

 The $d_{xy}$ pairing symmetry is quite robust here. In fact, if we consider the system in a strong electron-electron correlation region in which a short AFM interaction can be produced through the superexchange mechanism. We can easily argue that the pairing symmetry in this limit is still $d_{xy}$ based on the Hu-Ding principle\cite{hu_local_2012} which states the pairing symmetry is selected by the momentum space form factor of AFM exchange couplings that produce the largest weight on Fermi surfaces. In this case, we would expect the gap function is proportional to $sink_xsink_y$, in which the nodal points at the BZ boundary in the above RPA calculations will not appear.

\section{Discussion and Summary}
\begin{figure}[h]
\centerline{\includegraphics[width=0.5\textwidth]{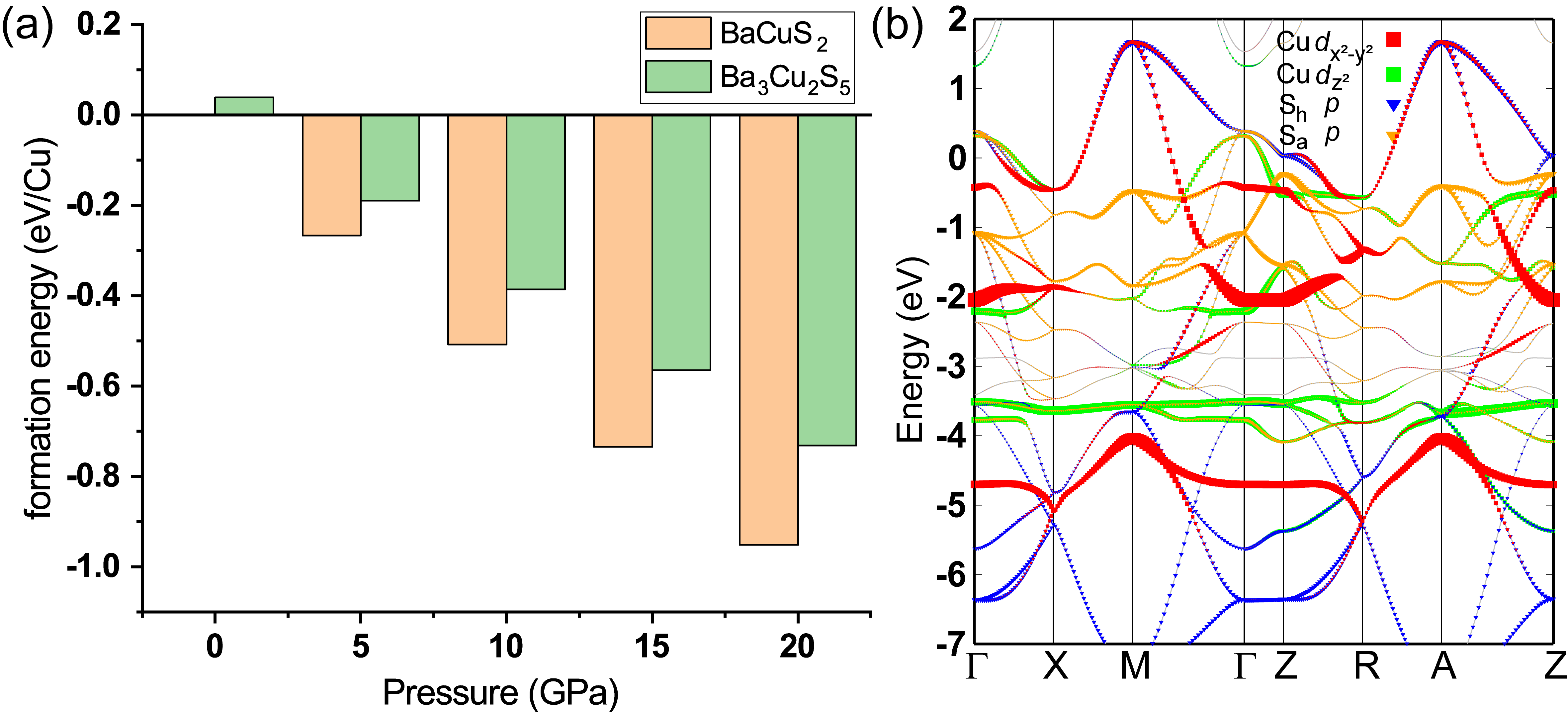}} \caption{(a) Formation energies of \ce{BaCuS2} and \ce{Ba3Cu2S5} under different pressures. (b) The band structure of \ce{BaCuS2} with an external pressure $P=$ 20 GPa from DFT calculation. The sizes of dots represent the weights of the projection.
 \label{pressure} }
\end{figure}

\begin{table}
\caption{\label{opt_volume}%
The optimized volumes of BaCuS$_2$, \ce{Ba3Cu2S5} and their decomposition phases under different pressures. All the data are normalized according to the formula. The unit of volume is $\AA^3$ here.}
\begin{ruledtabular}
\begin{tabular}{cccccc}
& 0 GPa & 5 GPa & 10 GPa  &15 GPa& 20 GPa\\
 \colrule
 \ce{BaCuS2} & 92.07 & 85.29 & 80.50 & 76.73 & 73.78 \\
 BaS+CuS & 101.41& 93.42& 87.99& 83.89& 80.55 \\
 \ce{Ba3Cu2S5} & 253.46 &  234.47 &  220.87 &  210.59 & 202.20 \\
 3BaS+2CuS & 270.09 & 247.89 & 232.96 & 221.74 & 212.65 \\
 \end{tabular}
\end{ruledtabular}
\end{table}

In summary, we propose a new superconducting material \ce{BaCuS2}. By comparing the electronic structures with \ce{CaCuO2}, \ce{FeSe} and \ce{La3N3Ni2B2}, we find that \ce{BaCuS2} should be a moderately correlated electron system with strong $p-d$ hybridization. The calculations based conventional BCS electron-phonon coupling suggests that it is a standard $s$-wave superconductor with $T_c<4$ K, while the electron-electron correlation results in an unconventional $B_{2g}$-wave superconductor and possibly much higher $T_c$.

 The structure of the material is in a highly stable phase according to our theoretical calculations.  In particular, we find that the structure has much lower formation energy than other known structures under external pressure. We calculate the formation energy of \ce{BaCuS2} and its sister compound \ce{Ba3Cu2S5}\cite{ba3cu2s5} under different pressures, as shown in FIG.\ref{pressure}.(a). The volume of \ce{BaCuS2}/\ce{Ba3Cu2S5} is remarkably less than that of BaS+CuS/3BaS+2CuS, as shown in TABLE.\ref{opt_volume}.  The main electronic physics does not vary much under pressure as shown in FiG.\ref{pressure}.(b), in which the band structure of \ce{BaCuS2} under 20 GPa is plotted. Therefore, it is promising that \ce{BaCuS2} can be synthesized in future experiments and studying its superconductivity can help us reveal the relationship between conventional and unconventional superconducting mechanisms.

{\it Acknowledgement:}We thank the useful discussions with Jianfeng Zhang and Yuechao Wang. This work is supported by the Ministry of Science and Technology of China 973 program
(No. 2017YFA0303100), National
Science Foundation of China (Grant No. NSFC11888101), and the
Strategic Priority Research Program of CAS (Grant
No.XDB28000000).

\bibliography{main}

\begin{thebibliography}{60}
\bibitem [{see()}]{see_appendix}%
  \BibitemOpen
  \href@noop {} {}\bibinfo {note} {Please find more details in our
  appendix.}\BibitemShut {Stop}%
\bibitem [{NOA()}]{NOAFM}%
  \BibitemOpen
  \href@noop {} {}\bibinfo {note} {In our DFT+$U$ calculations,
  antiferromagnetism always quenches ($U=0 \sim 10 eV$) in
  \ce{BaCuS2}.}\BibitemShut {Stop}%
\bibitem [{ba3()}]{ba3cu2s5}%
  \BibitemOpen
  \href@noop {} {}\bibinfo {note} {\ce{Ba3Cu2S5} shares similar electronic
  physics with \ce{BaCuS2}, and \ce{Cu} is also in square pyramidal
  coordination. Please find more details in our appendix.}\BibitemShut {Stop}%
\end{thebibliography}

\clearpage
\appendix
\renewcommand{\thetable}{S\arabic{table}}
\setcounter{table}{0}
\renewcommand{\thefigure}{S\arabic{figure}}
\setcounter{figure}{0}
\section{Computational methods}
Our electronic structure calculations employ the Vienna ab initio simulation package (VASP) code\cite{kresse1996} with the projector augmented wave (PAW) method\cite{Joubert1999}. The Perdew-Burke-Ernzerhof (PBE)\cite{perdew1996} exchange-correlation functional is used in our calculations. The kinetic energy cutoff is set to be 600 eV for the expanding the wave functions into a plane-wave basis in VASP calcuations. For body centered tetragonal \ce{La3N3Ni2B2} and \ce{Ba3Cu2S5}, we employ their primitive cells to perform calculations.
In the calculations of the formation energy, the energy convergence criterion is $10^{-6}$ eV and the force convergence criterion is 0.01 eV/\AA.  The $\Gamma$-centered \textbf{k}-meshes are $16\times16\times8$, $6\times6\times6$, $16\times16\times16$, $18\times18\times4$, $18\times18\times22$,  $20\times20\times12$ and $8\times8\times8$ for \ce{BaCuS2}, \ce{Ba3Cu2S5}, BaS,  CuS, \ce{CaCuO2}, FeSe and \ce{La3N3Ni2B2}, respectively.

We employ Wannier90\cite{mostofi2008wannier90,Marzari2012} to calculate maximally localized Wannier functions in \ce{BaCuS2}, \ce{CaCuO2}, FeSe and \ce{La3N3Ni2B2}. In the calculations of the $d$-$p$ models, the initial projectors are transition metal atoms' $d$-orbitals and anions' $p$-orbitals in \ce{BaCuS2}, \ce{CaCuO2} and FeSe. For \ce{La3N3Ni2B2}, the Ni($d$)-B($p$) valence manifold strongly entangles with other bands, so La's $d$-orbitals and B's $s$-orbitals are added in its initial projectors to reproduce DFT-calculated band structures. In the calculation of the $d$-$p$ $\sigma^*$ MLWFs, the initial projectors are Cu's $d_{x^2-y^2}+d_{z^2}$ orbitals in \ce{BaCuO2} and Cu's $d_{x^2-y^2}$ orbital in \ce{CaCuO2}, respectively.

We employ EPW package\cite{ponce2016epw} to calculate the electron-phonon coupling properties of \ce{BaCuS2}. The MLWFs are calculated by Wannier90\cite{mostofi2008wannier90,Marzari2012} interfacing with Quantum ESPRESSO\cite{giannozzi2009quantum}. We take the $16\times16\times8$ \textbf{k}-mesh and $4\times4\times2$ \textbf{q}-mesh as coarse grids and then interpolate to the $64\times64\times32$ \textbf{k}-mesh and $8\times8\times4$ \textbf{q}-mesh. The kinetic energy cutoff is set to 80 Ry. The Gaussian smearing method with the width of 0.005 Ry is used for the Fermi surface broadening. The energy convergence criterion is $10^{-12}$ eV. In the highly accurate structural optimization, the lattice constants and atomic coordinates are relaxed and the force convergence criterion is 0.000001 Ry/Bohr. The exchange-correlation functional is also PBE and the norm-conserving SG15 pseudopotentials are used\cite{Normconserving,hamann2013ONCV,schlipf2015sg15}.

\section{electron-phonon properties of B\lowercase{a}C\lowercase{u}S$_2$}
\begin{figure}[h]
\centerline{\includegraphics[width=0.5\textwidth]{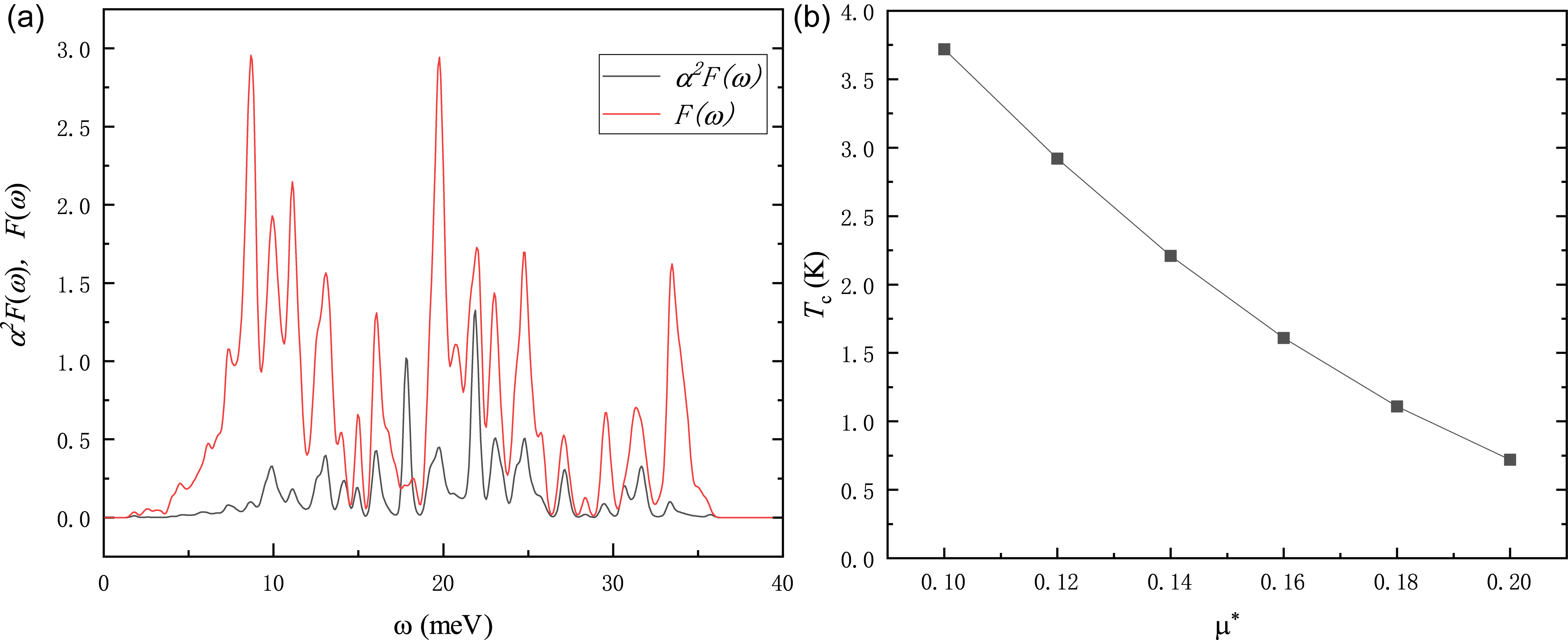}} \caption{(a) Eliashberg spectral function $\alpha^2F(\omega)$ (red line) and Phonon density of states $F(\omega)$ (black line) for \ce{BaCuS2}. (b) Evaluated $T_c$ as a function of $\mu^*$ for \ce{BaCuS2}.
 \label{elph} }
\end{figure}

The phonon density of states $F(\omega)$ and the corresponding Eliashberg spectral function $\alpha^2F(\omega)$ are plotted in FIG.\ref{elph}.(a). By intergating $\alpha^2F(\omega)$, we get a moderate EPC strength $\lambda=0.59$. We estimate the superconducting transition temperature $T_c$ with the McMillan-Allen-Dynes formula\cite{allen1972neutron,allen1975transition},
\begin{equation}T_{c}=\frac{\omega_{\log }}{1.2} \exp \left[\frac{-1.04(1+\lambda)}{\lambda\left(1-0.62 \mu^{*}\right)-\mu^{*}}\right],\end{equation}
where $\mu^*$ is the effective screened Coulomb repulsion constant and the
logarithmic average of the Eliashberg spectral function $\omega_{\log}$ is denfined as
\begin{equation}\omega_{\log }=\exp \left[\frac{2}{\lambda} \int \frac{d \omega}{\omega} \alpha^{2} F(\omega) \ln (\omega)\right].\end{equation} As $\mu^*$ is an input parameter, we plot $T_c$ as a function of $\mu^*$ in FIG.\ref{elph}.(b). The phonon-mediated $T_c$ for \ce{BaCuS2} should be less than 4 K.

\section{B\lowercase{a}$_3$C\lowercase{u}$_2$S$_5$: separation by three rock salt-type BaS layers}
\begin{figure}[h]
\centerline{\includegraphics[width=0.5\textwidth]{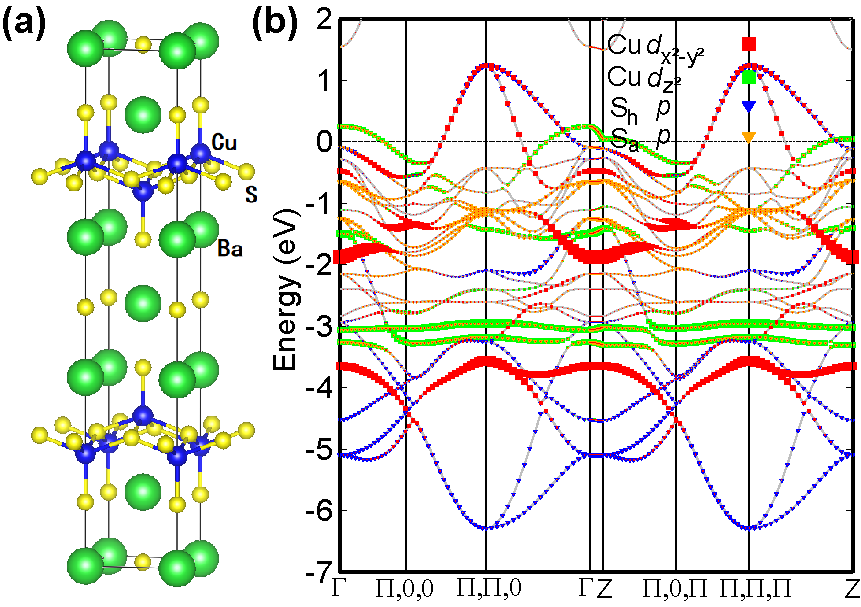}} \caption{(a) The crystal structure of \ce{Ba3Cu2S5}. (b) The band structure of \ce{BaCuS2} with its primitive cell from DFT calculation. The sizes of dots represent the weights of the projection. Here S$_a$ represents the apical S atoms while S$_h$ represents the horizontal S atoms. The choice of the $\mathbf{k}$-path is same as the literature's\cite{Pickett1995La3N3Ni2B2}.
 \label{ba3cu2s5} }
\end{figure}

As shown in FIG.\ref{ba3cu2s5}, the crystal structure of \ce{Ba3Cu2S5} is similar to that of \ce{BaCuS2}: The inverse $\alpha$-PbO-type \ce{Cu2S2} layer is separated by 3 rock salt-type BaS layers in \ce{Ba3Cu2S5} but separated by 2 BaS layers in \ce{BaCuS2} (\ce{Ba2Cu2S4}). It also shares a similar electronic structure with \ce{BaCuS2}, as shown in FIG.\ref{ba3cu2s5}.(b). \ce{Ba3Cu2S5} is not thermodynamically stable, but it is possible to synthesized \ce{Ba3Cu2S5} under external pressure due to Cu's five-coordination, as shown in FIG.\ref{pressure}.(a).

\section{Wannierization projected by \lowercase{$d$-}orbitals and \lowercase{$p$-orbitals}}
\begin{figure}[h]
\centerline{\includegraphics[width=0.5\textwidth]{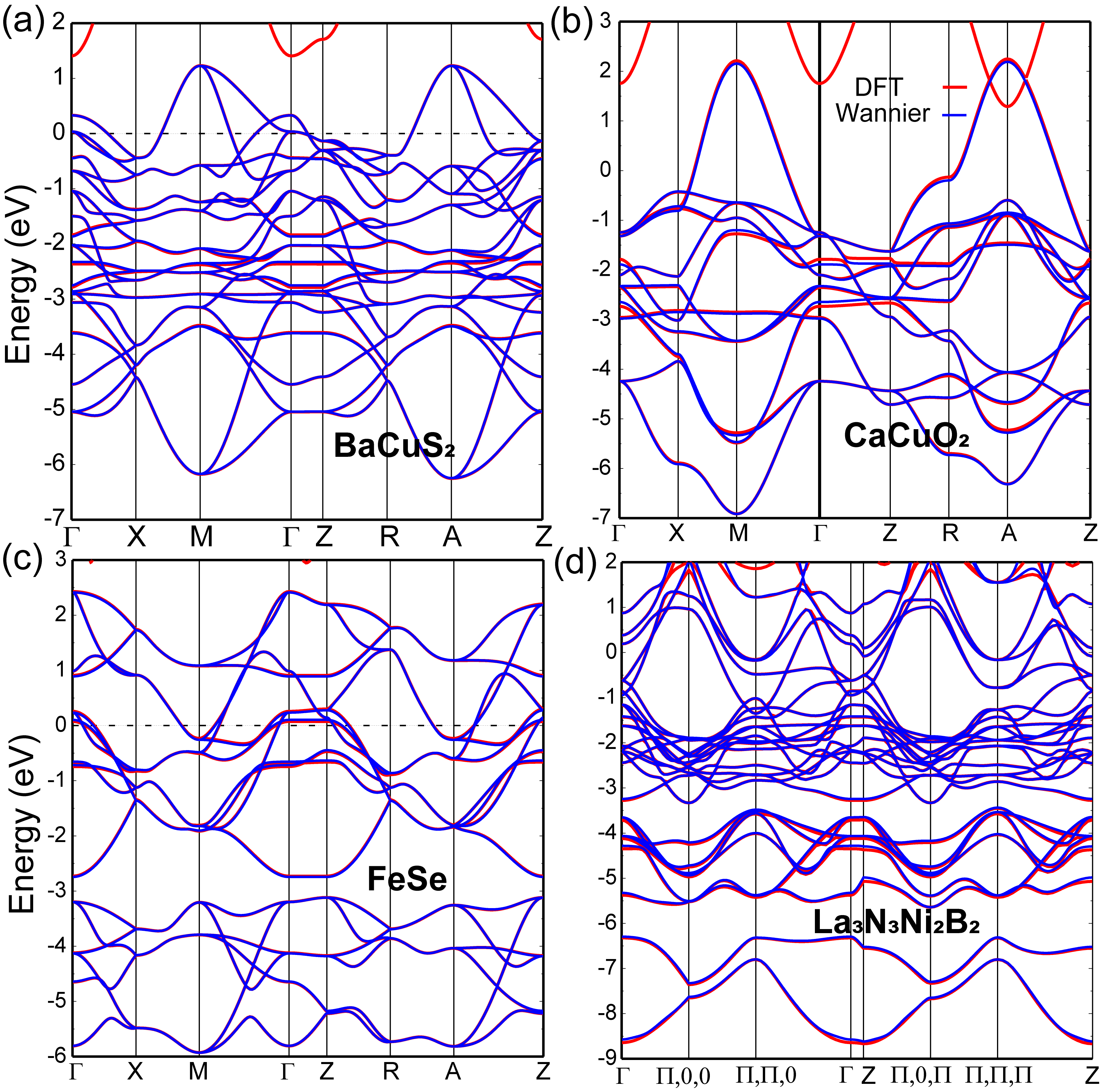}} \caption{The band structures of (a) \ce{BaCuS2}, (b) \ce{CaCuO2}, (c) \ce{FeSe} and (d) \ce{La3N3Ni2B2}. The red/blue lines represent DFT/Wannierization results, respectively. The choice of the $\mathbf{k}$-path in (d) is same as the literature's\cite{Pickett1995La3N3Ni2B2}.
 \label{dp_wan} }
\end{figure}

Our Wannierization results successfully reproduce DFT-calculated band structures, as shown in FIG.\ref{dp_wan}. The relevant representative hopping parameters and on-site energies are listed in TABLE.\ref{hops}. Here we use the conventional notations of the local crystal field coordinations.
\begin{table}[ht]
\caption{\label{hops}%
The hopping parameters and on-site energies for \ce{BaCuS2}, \ce{CaCuO2}, FeSe and \ce{La3N3Ni2B2}. Here S$_a$ represents the apical S atoms while S$_h$ represents the horizontal S atoms.}
\begin{ruledtabular}
\begin{tabular}{cc}
 \ce{BaCuS2}&  \\
 \colrule
 $\varepsilon_{Cu,d_{z2}}$ & -2.21\\
 $\varepsilon_{Cu,d_{x^2-y^2}}$ & -2.23\\
 $\varepsilon_{Cu,d_{xz/yz}}$ & -2.32\\
 $\varepsilon_{Cu,d_{xy}}$ & -2.48\\
 $\varepsilon_{S_h,p_{z}}$ & -2.58\\
 $\varepsilon_{S_h,p_{x/y}}$ & -3.08\\
 $\varepsilon_{S_a,p_{z}}$ & -2.17\\
 $\varepsilon_{S_a,p_{x/y}}$ & -1.40\\
 $|t_{{Cu,d_{x^2-y^2}}-{S_h,p_{z}}}|$ & 0.37\\
 $|t_{{Cu,d_{x^2-y^2}}-{S_h,p_{x/y}}}|$ & 0.69\\
 $|t_{{Cu,d_{z^2}}-{S_h,p_{z}}}|$ & 0.39\\
 $|t_{{Cu,d_{z^2}}-{S_h,p_{x/y}}}|$ & 0.11\\
 $|t_{{Cu,d_{z^2}}-{S_a,p_{z}}}|$ & 0.82\\
 $|t_{{S_a,p_{x}}-{S_a,p_x}}|$ & 0.09\\
 \colrule
  \ce{CaCuO2}&  \\
 \colrule
  $\varepsilon_{Cu,d_{z2}}$ & -2.42\\
 $\varepsilon_{Cu,d_{x^2-y^2}}$ & -1.92\\
  $\varepsilon_{O,p_{z}}$ & -2.58\\
 $\varepsilon_{O,p_{x/y}}$ & -3.83\\
 $|t_{{Cu,d_{x^2-y^2}}-{O,p_{x/y}}}|$ & 1.24\\
 \colrule
  FeSe &  \\
 \colrule
 $\varepsilon_{Fe,d_{x^2-y^2}}$ & -0.88\\
  $\varepsilon_{Fe,d_{xz/yz}}$ & -0.78\\
  $\varepsilon_{Se,p_{z}}$ & -3.07\\
 $\varepsilon_{Se,p_{x/y}}$ & -3.09\\
 $|t_{{Fe,d_{x^2-y^2}}-{Se,p_{x/y}}}|$ & 0.25\\
 $|t_{{Fe,d_{x^2-y^2}}-{Se,p_{z}}}|$ & 0.72\\
 $|t_{{Fe,d_{xz/yz}}-{Se,p_{x/y}}}|$ & 1.00\\
 $|t_{{Fe,d_{x^2-y^2}}-{Se,p_{z}}}|$ & 0.16\\
 \colrule
 \ce{La3N3Ni2B2} &  \\
 \colrule
 $\varepsilon_{Ni,d_{z^2}}$ & -2.11\\
  $\varepsilon_{Ni,d_{x^2-y^2}}$ & -2.26\\
  $\varepsilon_{Ni,d_{xz/yz}}$ & -2.11\\
  $\varepsilon_{Ni,d_{xy}}$ & -2.24\\
  $\varepsilon_{B,s}$ & 0.23\\
  $\varepsilon_{B,p_{z}}$ & 2.17\\
  $\varepsilon_{B,p_{x/y}}$ & 0.28\\
  $|t_{{Ni,d_{x^2-y^2}}-{Ni,p_{x/y}}}|$ & 0.55\\
 $|t_{{Ni,d_{x^2-y^2}}-{Ni,p_{z}}}|$ & 0.78\\
 $|t_{{Ni,d_{xz/yz}}-{Ni,p_{x/y}}}|$ & 0.85\\
 $|t_{{Ni,d_{x^2-y^2}}-{Ni,p_{z}}}|$ & 0.27\\
 \end{tabular}
\end{ruledtabular}
\end{table}

\clearpage

\section{Wannierization of the \lowercase{$d$-$p$ $\sigma^*$}-bonding bands and the effective tight-binding model}
\begin{figure}[htb]
\centerline{\includegraphics[width=0.5\textwidth]{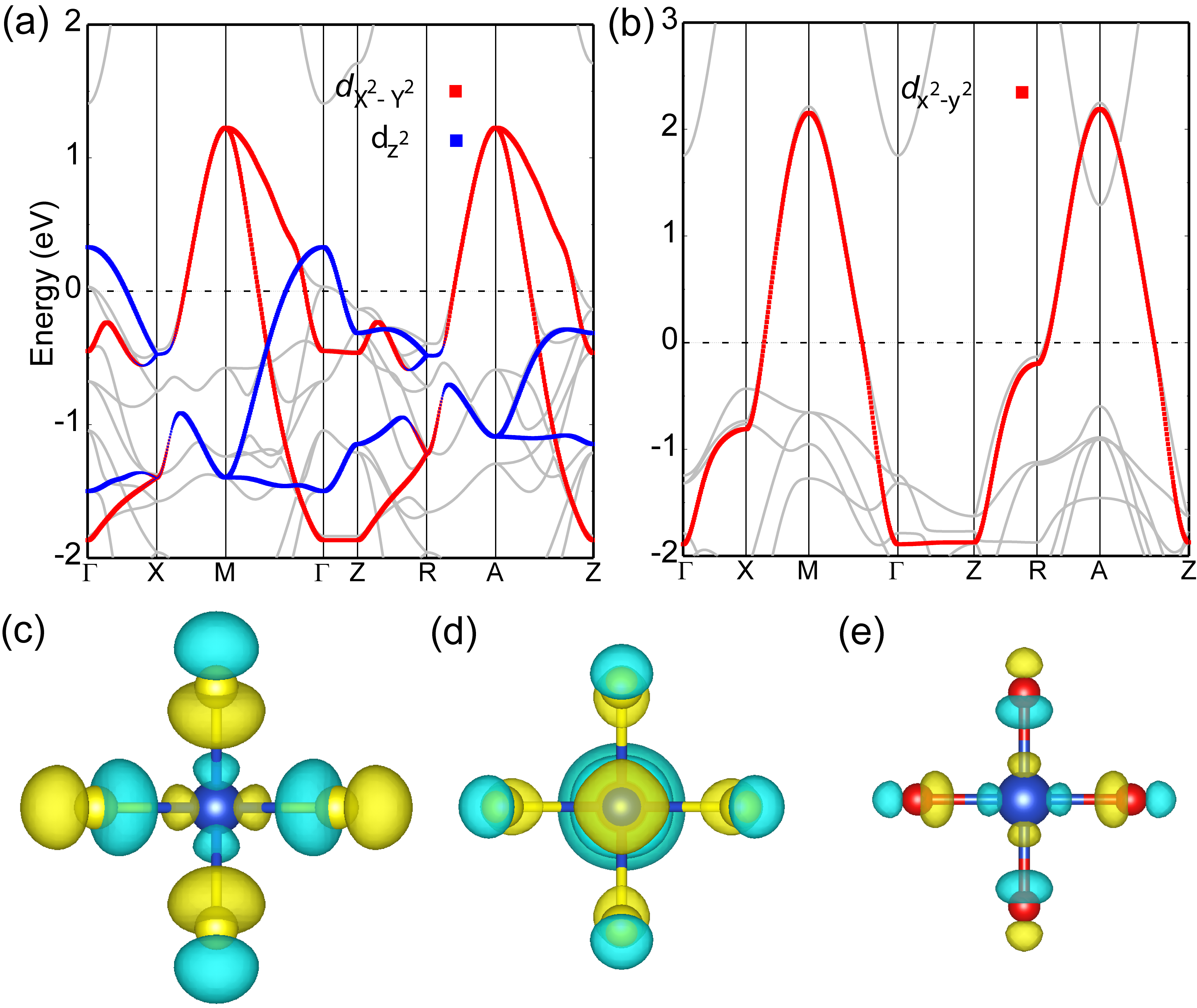}} \caption{(a-b) The band structures of (a) \ce{BaCuS2} and (b) \ce{CaCuO2} calculated by DFT (gray lines) and Wannierizaiton (red/blue dots). The sizes of dots represent the weights of the projection of the $d$-$p$ $\sigma^*$ Wannier functions. (c-d) The isosurface of (c) the $d_{X^2-Y^2}$-like MLWF and (d) the $d_{z^2}$-like MLWF in \ce{BaCuS2}. (e) The isosurface of the $d_{x^2-y^2}$-like MLWF in \ce{CaCuO2}.
 \label{sigma*_wan} }
\end{figure}

As mentioned above, the in-plane $d$-$p$ $\sigma^*$-bonding bands are isolated around the Fermi surface. In order to construct the effective minimal model to describe the in-plane electronic physics in \ce{BaCuS2}, we downfold the full $d-p$ model into an effective minimal model\cite{gu2020substantial} by only Wannierizing the $d_{X^2-Y^2}$-like and $d_{z^2}$-like MLWFs in \ce{BaCuS2} with a smaller energy window. Our Wannierization results capture the main characters of \ce{BaCuS2}'s electronic structure, as shown in FIG.\ref{sigma*_wan}.(a). This is an analogy to the Zhang-Rice singlet in cuprates\cite{zhang_Rice1988}, so we also calculate the $d_{X^2-Y^2}$-like MLWF in \ce{CaCuO2} for comparison, as shown in FIG.\ref{sigma*_wan}.(b).

We construct the effective tight-binding (TB) model in the basis of $d_{X^2-Y^2}$ orbital and $d_{z^2}$ orbital to describe the in-plane electronic physics. Since there are two Cu atoms in one unit cell, the TB model can be written as a $4\times4$ Hermitian matrix:
\begin{widetext}
\begin{eqnarray}\label{H44}
H_{11}&=&H_{33}=\varepsilon_1+2t_{11}^x(cos(k_x)+cos(k_y))+2t_{11}^{xx}(cos(2k_x)+cos(2k_y))+4t_{11}^{xxyy}(cos(k_x)cos(k_y)),\nonumber \\
H_{12}&=&H_{34}=2t_{12}^x(cos(k_x)-cos(k_y))+2t_{12}^{xx}(cos(2k_x)-cos(2k_y)),\nonumber \\
H_{13}&=&4t_{13}^{xy}cos(k_x/2)cos(k_y/2)+4t_{13}^{xxy}(cos(k_x/2)*cos(3k_y/2)+cos(3k_x/2)cos(k_y/2)),\nonumber \\
H_{14}&=&H_{23}=4t_{14}^{xxy}(cos(3k_x/2)cos(k_y/2)-cos(k_x/2)cos(3k_y/2)),\nonumber \\
H_{22}&=&H_{44}=\varepsilon_2+2t_{22}^x(cos(k_x)+cos(k_y))+2t_{22}^{xx}(cos(2k_x)+cos(2k_y))+4t_{22}^{xxyy}(cos(k_x)cos(k_y)),\nonumber \\
H_{24}&=&4t_{24}^{xy}cos(k_x/2)cos(k_y/2)+4t_{24}^{xxy}(cos(k_x/2)*cos(3k_y/2)+cos(3k_x/2)cos(k_y/2)). \nonumber\\
\end{eqnarray}
\end{widetext}
The hopping parameters are truncated to the fifth-nearest-neighbour site. We get hopping parameters and on-site energies by fitting to the Wannierization result in $k_z=0$ plane, as shown in FIG.\ref{TB_band}. The corresponding parameters and their notations are listed in TABLE.\ref{TB_hops}. The major hopping parameter is $t_{11}^x$, the intra-orbital hopping between two SNN $d_{X^2-Y^2}$ orbital, which is in the same energy scale with the dominating intra-orbital hopping between two NN $d_{x^2-y^2}$ orbital in cuprates ($t^{NN}_{d_{x^2-y^2}}$ is about -0.47 in \ce{CaCuO2}).

As mentioned in our main text, we can transfer the $4\times4$ TB model into a block-diagonalized matrix with using the glide symmetry:
\begin{equation}
H_{eff}(\textbf{k})=
    \begin{pmatrix}
        H_k & 0 \\
        0 & H_{k+Q}\\
    \end{pmatrix},
\end{equation}
here $H_k$ is the effective two-band model in our main text and $Q=(\pi,\pi)$. The explict form of $H_k$ is
\begin{eqnarray}
H_k =
    \begin{pmatrix}
        H_{11}+H_{31} & H_{12}+H_{32} \\
        H_{21}+H_{41} & H_{22}+H_{42}\\
    \end{pmatrix},
\end{eqnarray}
where $H_{\alpha\beta}$ are matrix elements in Eq.\ref{H44}.

Visually, we plot these $d$-$p$ $\sigma^*$ Wannier functions in \ce{BaCuS2} and \ce{CaCuO2}, as shown in FIG.\ref{sigma*_wan}.(c-e). These Wannier functions are composed of Cu's $d$-orbitals and coordinated S/O's $p$-orbitals symmetrically. As the isovalues of isosurfaces in FIG.\ref{sigma*_wan}.(c-e) are same,  the $d$-$p$ $\sigma^*$-bonding bands are more delocalized in \ce{BaCuS2} than that in \ce{CaCuO2}. As a result, the correlation strength in \ce{BaCuS2} should be weaker.

 \begin{figure}[h]
\centerline{\includegraphics[width=0.5\textwidth]{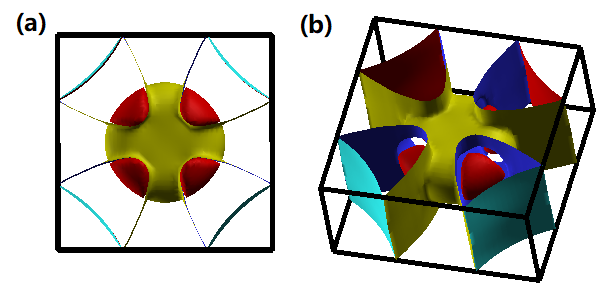}} \caption{The Fermi surfaces of \ce{BaCuS2} by Wannier fitting with 4 MLWFs from (a) top view and (b) oblique view.
 \label{twoband_FS} }
\end{figure}

 \begin{figure}[h]
\centerline{\includegraphics[width=0.5\textwidth]{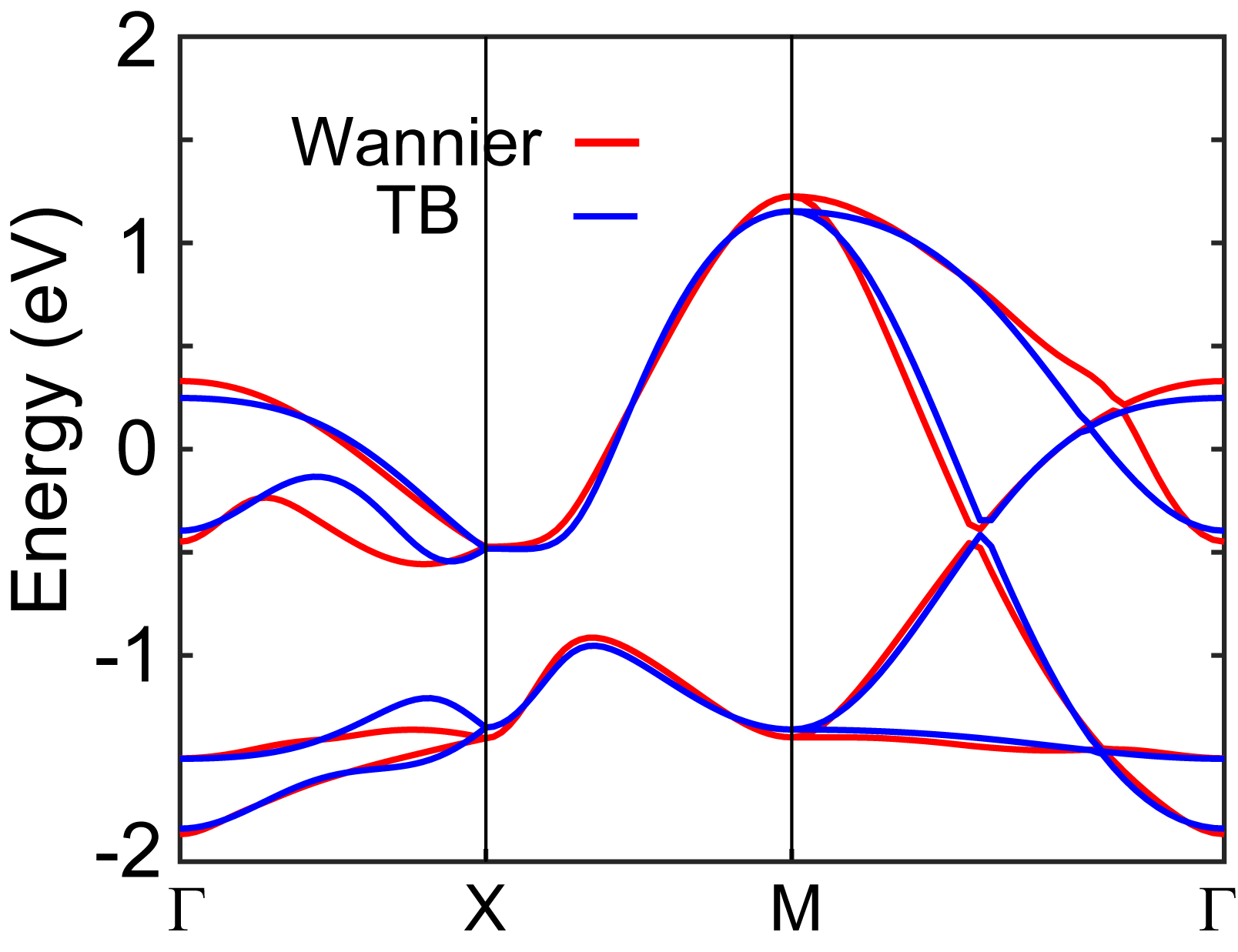}} \caption{Comparision of the band structures of \ce{BaCuS2} by Wannierizaiton (red lines) and fitted TB model (blue lines).
 \label{TB_band} }
\end{figure}

\begin{table}
\caption{\label{TB_hops} The hopping parameters and on-site energies of in-plane TB model for \ce{BaCuS2}. The energy unit is eV. Here superscript $x$ labels the hopping between two second-nearest-neighbour (SNN) sites along $X$ direction, superscript $xx$ labels the hopping between two forth-nearest-neighbour sites along $X$ direction, superscript $xy$ labels the hopping between two nearest-neighbour (NN) sites along $Y=X$ direction, superscript $xxyy$ labels the hopping between two third-nearest-neighbour (TNN) sites along $Y=X$ direction, superscript $xxy$ labels the hopping between two fifth-nearest-neighbour sites along $Y=X/3$ direction; subscript 1-4 represent \ce{Cu_A}'s $d_{X^2-Y^2}$ orbital, \ce{Cu_A}'s $d_{z^2}$ orbital, \ce{Cu_B}'s $d_{X^2-Y^2}$ orbital and \ce{Cu_B}'s $d_{z^2}$ orbital, respectively.}
\begin{ruledtabular}
\begin{tabular}{ccccc}
$\varepsilon_1$ & $\varepsilon_2$ & $t_{11}^x$ & $t_{11}^{xx}$  & $t_{11}^{xxyy}$  \\
  -0.31 & -0.82 &  -0.28  &  -0.07  & 0.15  \\
  \hline
$t_{22}^{x}$  & $t_{22}^{xx}$  & $t_{22}^{xxyy}$ & $t_{12}^{x}$  & $t_{12}^{xx}$\\
   0.09  &  0.003  & -0.05  & -0.08 & 0.01 \\
  \hline
  $t_{13}^{xy}$  & $t_{13}^{xxy}$  & $t_{24}^{xy}$ & $t_{24}^{xxy}$  & $t_{14}^{xxy}$\\
   0.25  & -0.03 & -0.26  &  0.02  & -0.01 \\
\end{tabular}
\end{ruledtabular}
\end{table}

\clearpage
\section{$U/J$ parameters calculated by Local screened Coulomb correction (LSCC) approach}
The $U/J$ parameters represent the correlation strength in DFT$+U$ calculations and are often chosen empirically. Here we employ the first-principle LSCC approach\cite{wang2019local} to calculate the $U/J$ parameters in layered transition metal compounds \ce{CaCuO2}, \ce{FeSe}, \ce{La3N3Ni2B2}, \ce{BaNiS2} and \ce{BaCuS2}. In LSCC method, the local Coulomb interactions are calculated by using the Yukawa potential, so the $U/J$ should decrease when the system becomes more metallic. Since the $U/J$ is strongly dependent on the muffin-tin radium $R_{MT}$, we should only compare the $U/J$ with the same pseudopotential. As shown in TABLE.\ref{LSCC}, the $U/J$ is larger when AFM order exists in \ce{CaCuO2}/\ce{FeSe}. The $U/J$ in \ce{BaNiS2} is larger that in \ce{La3N3Ni2B2} because the correlation effect are non-negligible in \ce{BaNiS2}\cite{klein2018banis2,santos2016anomalous_banis2} and \ce{La3N3Ni2B2} is a typical metal. From this point of view, our results also 
demonstrate that the correlation in \ce{BaCuS2} is weaker than in cuprate \ce{CaCuO2}.

\begin{table}[h]
\caption{\label{LSCC}%
The $U/J$ parameters and moments calculated by LSCC method. }
\begin{ruledtabular}
\begin{tabular}{cccc}
 \colrule
LSCC & U (eV) & J (eV) & moment ($\mu_B$) \\
\ce{CaCuO2}(AFM) & 5.78 & 1.16 & 0.478 \\
\ce{CaCuO2}(NM) & 5.74 & 1.16 & 0 \\
FeSe(CAFM) & 4.88 & 0.91 & 3.05 \\
FeSe(NM) & 4.75 & 0.89 & 0 \\
\ce{La3N3Ni2B2} & 5.49 & 0.99 & 0 \\
\ce{BaNiS2} & 5.61 & 1.01 & 0 \\
\ce{BaCuS2} & 5.7 & 1.15 & 0 \\
 \end{tabular}
\end{ruledtabular}
\end{table}

\section{method of RPA calculation}
In this section, we explain the formalism of the multiorbital RPA approach\cite{berk1966effect,scalapino1986d,graser2009near,kemper2010sensitivity,xxwu2015triplet}, adopted in the main text. The multi-orbital susceptibility is defined as,
\begin{align}
\chi_{l_1l_2l_3l_4}(\bm{q},\tau)=&\frac{1}{N}\sum_{\bm{k}\bm{k}'}\langle T_{\tau}c^{\dag}_{l_3\sigma}(\bm{k}+\bm{q},\tau) \\ \nonumber
&c_{l_4\sigma}(\bm{k},\tau)c^{\dag}_{l_2\sigma'}(\bm{k}'-\bm{q},0)c_{l_1\sigma'}(\bm{k}',0) \rangle .
\end{align}
In momentum-frequency space, the multi-orbital bare susceptibility is given by
\begin{small}
\begin{align}
\chi^{0}_{l_{1}l_{2}l_{3}l_{4}}(\bm{q},i\omega_{n})=&-\frac{1}{N}\sum_{\bm{k}\mu\nu}a^{l_{4}}_{\mu}(\bm{k})a^{l_{2}*}_{\mu}(\bm{k})a^{l_{1}}_{\nu}(\bm{k}+\bm{q}) \\ \nonumber
&a^{l_{3}*}_{\nu}(\bm{k}+\bm{q})\frac{n_{F}(E_{\mu}(\bm{k}))-n_{F}(E_{\nu}(\bm{k}+\bm{q}))}{i\omega_{n}+E_{\mu}(\bm{k})-E_{\nu}(\bm{k}+\bm{q})},
\end{align}
\end{small}
where $\mu$ and $\nu$ are the band indices, $n_{F}$ is the usual Fermi distribution, $l_{i}$ $(i=1,2,3,4)$ are the orbital indices, $a^{l_{i}}_{\mu}(k)$ is the $l_{i}$ orbital component of the eigenvector for band $\mu$ resulting from the diagonalization of the tight-binding Hamiltonian $H_{0}$ and $E_{\mu}(\bm{k})$ is the corresponding eigenvalue. With interactions, the RPA spin and charge susceptibilities are given by
\begin{equation}
\begin{aligned}
\chi^{RPA}_{s}(\bm{q})=\chi^{0}(\bm{q})[1-\bar{U}^{s}\chi^{0}(\bm{q})]^{-1},\\
\chi^{RPA}_{c}(\bm{q})=\chi^{0}(\bm{q})[1+\bar{U}^{c}\chi^{0}(\bm{q})]^{-1},
\end{aligned}
\end{equation}
where $\bar{U}^{s}$ ($\bar{U}^{c}$) is the spin (charge) interaction matrix,
\begin{eqnarray}
\bar{U}^s_{l_1l_2l_3l_4}(\bm{q})&=&
\begin{cases}
U    & l_1=l_2=l_3=l_4,\\
U'     & l_1=l_3\neq l_2=l_4,\\
J   & l_1=l_2\neq l_3=l_4,\\
J'   & l_1=l_4\neq l_2=l_3,\\
\end{cases}\\
\label{EQ:Us}
\bar{U}^c_{l_1l_2l_3l_4}(\bm{q})&=&
\begin{cases}
U   & l_1=l_2=l_3=l_4,\\
-U'+2J    & l_1=l_3\neq l_2=l_4,\\
2U'-J    & l_1=l_2\neq l_3=l_4,\\
J'   & l_1=l_4\neq l_2=l_3,\\
\end{cases}.
\label{EQ:Uc1}
\end{eqnarray}
In the main text, we plot the largest eigenvalues of the susceptibility matrix $\chi^{0}_{l_{1}l_{1}l_{2}l_{2}}(\bm{q},0)$ and $\chi^{RPA}_{s,l_{1}l_{1}l_{2}l_{2}}(\bm{q},0)$. Within RPA approximation, the effective Cooper scattering interaction on Fermi surfaces is,
\begin{align}
\Gamma_{ij}(\bm{k},\bm{k}')=&\sum_{l_{1}l_{2}l_{3}l_{4}}a^{l_{2},\ast}_{\emph{v}_{i}}(\bm{k})a^{l_{3},\ast}_{\emph{v}_{i}}(-\bm{k}) \\ \nonumber
&\emph{Re}\bigg[\Gamma_{l_{1}l_{2}l_{3}l_{4}}(\bm{k},\bm{k}',\omega=0)\bigg]a^{l_{1}}_{\emph{v}_{j}}(\bm{k}')a^{l_{4}}_{\emph{v}_{j}}(-\bm{k}'),
\end{align}
where the momenta $\bm{k}$ and $\bm{k}'$ is restricted to different FSs with $\bm{k}\in C_{i}$ and $\bm{k}'\in C_{j}$. The  orbital vertex function $\Gamma_{l_{1}l_{2}l_{3}l_{4}}$ in spin singlet channel\cite{takimoto2004strong,kubo2007pairing} is
\begin{small}
\begin{align}
\Gamma^{S}_{l_{1}l_{2}l_{3}l_{4}}(\bm{k},\bm{k}',\omega)=&\bigg[\frac{3}{2}\bar{U}^{s}\chi^{RPA}_{s}(\bm{k}-\bm{k}',\omega)\bar{U}^{s}+\frac{1}{2}\bar{U}^{s} \\ \nonumber
&-\frac{1}{2}\bar{U}^{c}\chi^{RPA}_{c}(\bm{k}-\bm{k}',\omega)\bar{U}^{c}+\frac{1}{2}\bar{U}^{c}\bigg]_{l_{1}l_{2}l_{3}l_{4}},
\end{align}
\end{small}
where $\chi^{RPA}_{s}$ and $\chi^{RPA}_{c}$ are the RPA spin and charge susceptibility, respectively. The pairing strength functional for a specific pairing state is given by,
\begin{equation}
\begin{aligned}
\lambda\big[\emph{g}(\bm{k})\big]=-\frac{\sum_{ij}\oint_{C_{i}}\frac{d\bm{k}_{\|}}{\emph{v}_{\emph{F}}(\bm{k})}\oint_{C_{j}}\frac{d\bm{k}'_{\|}}{\emph{v}_{\emph{F}}(\bm{k}')}\emph{g}(\bm{k})\Gamma_{ij}(\bm{k},\bm{k}')\emph{g}(\bm{k}')}{(2\pi)^{2}\sum_{i}\oint_{C_{i}}\frac{d\bm{k}_{\|}}{\emph{v}_{\emph{F}}(\bm{k})}\big[\emph{g}(\bm{k})\big]^{2}},
\end{aligned}
\end{equation}
where $v_{F}(\bm{k})=|\nabla_{k}E_{i}(\bm{k})|$ is the Fermi velocity on a given Fermi surface sheet $C_{i}$. The pairing vertex function in spin singlet and triplet channels are symmetric and antisymmetric parts of the interaction, that is, $\Gamma^{S/T}_{ij}(\bm{k},\bm{k}')=\frac{1}{2}[\Gamma_{ij}(\bm{k},\bm{k}')\pm \Gamma_{ij}(\bm{k},-\bm{k}')]$.

\end{document}